\documentclass[peerreview,transmag,12pt,onecolumn]{IEEEtran}
\usepackage[T1]{fontenc}
\usepackage[utf8]{inputenc}
\usepackage{multirow}
\usepackage[tbtags]{amsmath}
\usepackage[cmintegrals]{newtxmath}
\usepackage{graphicx}
\usepackage{array}
\usepackage[numbers]{natbib}

\providecommand{\tabularnewline}{\\}

\usepackage{fixmath}
\usepackage{bm}
\usepackage{booktabs}
\usepackage{makecell}
\usepackage{xcolor}  
\usepackage{mciteplus}
\usepackage[english]{babel}
\usepackage{hyperref}
\newcommand\blfootnote[1]{%
  \begingroup
  \renewcommand\thefootnote{}\footnote{#1}%
  \addtocounter{footnote}{-1}%
  \endgroup
}
\newcommand*\V[1]{\bm{{#1}}}
\newcommand*\T[1]{\hat{#1}}
\newcommand*\diff{\mathrm{d}}
\newcommand*\imun{\mathrm{i}}
\newcommand*\tv{\mathrm{t}}
\newcommand*\grad{\operatorname{\mathbf{grad}}}
\newcommand*\curl{\operatorname{\mathbf{curl}}}
\newcommand*\divg{\operatorname{div}}
\newcommand*\gradt{\operatorname{\mathbf{grad}}_{\tv}}
\newcommand*\scurlt{\operatorname{curl}_{\tv}}
\newcommand*\vcurlt{\operatorname{\mathbf{curl}}_{\tv}}
\newcommand*\divgt{\operatorname{div}_{\tv}}
\newcommand*\modimpt{\bar{\nu}_{\tv}}
\newcommand*\modimptenst{\T{\bar{\nu}}_{\tv}}
\newcommand*\lctenst{\T{\epsilon}_{\tv}}
\newcommand*\Omegat{\Omega_{\tv}}
\newcommand*\Gammat{\Gamma_{\tv}}

\newcommand*\twocolbreak{%
\ifCLASSOPTIONtwocolumn
\\%
\fi
}

\begin{document}
\title{2D Fourier finite element formulation for magnetostatics in curvilinear coordinates with a symmetry direction}
\author{\IEEEauthorblockN{Christopher G. Albert\IEEEauthorrefmark{1,2}, Oszkár Bíró\IEEEauthorrefmark{3}, Patrick Lainer\IEEEauthorrefmark{2}}
\\
\IEEEauthorblockA{\IEEEauthorrefmark{1}Max-Planck-Institut für Plasmaphysik, Boltzmannstraße 2, 85748 Garching, Germany}
\\
\IEEEauthorblockA{Graz University of Technology -- Graz Center of Computational Engineering}
\IEEEauthorblockA{\IEEEauthorrefmark{2}Institute of Theoretical and Computational Physics, Petersgasse 16, 8010 Graz, Austria
}
\IEEEauthorblockA{\IEEEauthorrefmark{3}Institute of Fundamentals and Theory in Electrical Engineering, Inffeldgasse 18, 8010 Graz, Austria}}
\maketitle
\begin{abstract}
We present a numerical method for the solution of linear magnetostatic problems
in domains with a symmetry direction, including axial and translational
symmetry. The approach uses a Fourier series decomposition
of the vector potential formulation along the symmetry direction and
covers both, zeroth (non-oscillatory) and non-zero (oscillatory) harmonics.
For the latter it is possible to eliminate one component of the vector
potential resulting in a fully transverse vector potential orthogonal
to the transverse magnetic field. In addition to the Poisson-like
equation for the longitudinal component of the non-oscillatory problem,
a general curl-curl Helmholtz equation results for the transverse
problem covering both, non-oscillatory and oscillatory case. The derivation
is performed in the covariant formalism for curvilinear coordinates
with a tensorial permeability and symmetry restrictions on metric
and permeability tensor. The resulting variational forms are treated
by the usual nodal finite element method for the longitudinal problem
and by a two-dimensional edge element method for the transverse problem.
The numerical solution can be computed independently for each harmonic
which is favourable with regard to memory usage and parallelisation.
\end{abstract}

\section{Introduction}

\blfootnote{\\\\ \textbf{Preprint version: \today}\\\\}Two-dimensional
formulations of electrodynamic problems in translationally and axisymmetric
systems are commonly used for numerical treatment by the finite element
method~\citep{Bladel2007-Book,Jin2015-Book}. This includes simplified
models of three-dimensional systems, as the number of degrees of freedom
is substantially reduced compared to the full model. In the simplest
case all quantities are assumed to be independent of the symmetry
variable, thereby reducing the problem dimension by one. In the more
general Fourier finite element method also oscillatory field components
are treated by combination of a Fourier expansion in the symmetry
coordinate with a numerical solution for each individual harmonic
(see, e.g.,~\citep{Bernardi1999-Spectral,lacoste2000solution,Oh2015-2063}
for the treatment of Maxwell's equations). For linear problems the
Fourier finite element method has the advantage of trivial parallelisability
over individual harmonics in comparison to a 3D computation.

Here we treat the magnetostatic problem using the vector potential
\textbf{curl}-\textbf{curl} formulation in domains with a symmetry
direction, including in particular the translationally and axisymmetric
case. Due to the linearity of the \textbf{curl} operator, it is possible
to choose a different gauge for each Fourier harmonic and obtain a
valid vector potential as the total sum. A similar approach has been
taken for the divergence-free interpolation of a given magnetic field
in~\citep{Heyn2008-24005}. As a result, the oscillatory part of the \textbf{curl}-\textbf{curl} equation can
be treated as a a fully two-dimensional problem using a transverse vector
potential. The resulting equations can be viewed as a generalisation
of the transverse part of the axisymmetric problem and are in analogy
to the transverse formulation of waveguide models~\citep{Dillon94}.
The approach is implemented using the metric-free covariant formulation
of Maxwell's equations in curvilinear coordinates (see \citep{schouten1954tensor,Post1997-Formal}
for details or \citep{Gronwald2005-506219} for a concise introduction)
that is briefly restated in the notation used for further derivations.
Here linear constitutive relations contain the permeability tensor
represented by its covariant density components implicitly including
the influence from curvilinear coordinates via their respective metric
tensor. While being mathematically equivalent to the formulation via
differential forms the covariant notation remains closer to the traditional
formulation via differential operators in flat space. For convenience,
Table~\ref{tab:coordinates-2} lists the relevant designations for both approaches.

Explicit expressions with a scalar permeability are given for the
Cartesian case with $2\pi$-periodicity in $z$-direction and axisymmetric
systems in cylindrical coordinates. Application to other axisymmetric
systems such as spherical, toroidal or symmetry flux coordinates \citep{Dhaeseleer_BOOK_1991}
is straightforward by inserting the components of their respective
metric tensor. As long as the latter fulfils the necessary requirements,
even more exotic coordinate systems (e.g. helical \citep{Garavaglia1975-1})
should be realisable. The resulting variational problems are discretised
in coordinate space using a two-dimensional (Nédélec) edge finite
element formulation~\citep{Biro1999-391--405,brezzi2012mixed} of
lowest order for the transverse equations and first order Lagrange
elements for the longitudinal non-oscillatory Poisson-like equation. A convergence
study for benchmarking problems with existing analytical solutions
in an axisymmetric domain is given. 

\section{Derivation of the method}

The magnetostatic equations for magnetic field $\V{H}(\V{r})$,
magnetic flux density $\V{B}(\V{r})$, and current density
$\V{J}(\V{r})$ are
\begin{align}
\curl \V{H} &= \V{J}, \label{eq:magstat-1} \\
\divg \V{B} &= 0, \label{eq:divb-1}
\end{align}
where $\V{H}$ is related to $\V{B}$ via the constitutive relation
\begin{equation}
\V{H} = \T{\nu} \V{B} \label{eq:constitu}
\end{equation}
with local reluctivity (inverse permeability) tensor $\T{\nu}(\V{r}) = \T{\mu}(\V{r})^{-1}$.
Written in terms of a vector potential $\V{A}(\V{r})$ with
$\curl \V{A} = \V{B}$, such that Eq.~(\ref{eq:divb-1})
is automatically fulfilled, they are equivalent to the curl-curl equation,
\begin{align}
\curl (\T{\nu} \curl \V{A}) &= \V{J}, \label{eq:curlcurl-1}
\end{align}
as long as the domain is simply connected. 

\begin{figure}[bth]
\centering
\includegraphics[width=3.5in]{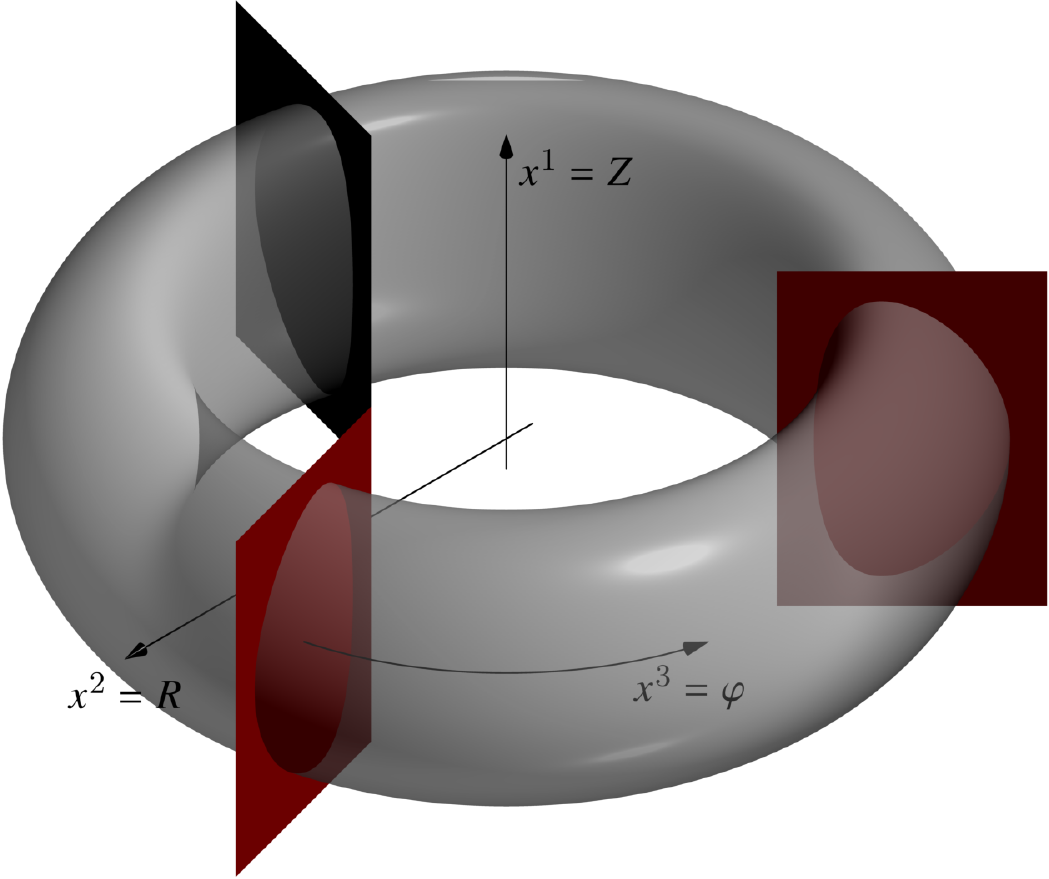}
\caption{Example of an axisymmetric domain with indication of the coordinate axes for $x^{1}, x^{2}, x^{3}$. Note that the indicated transverse cross-sections do not change their shape along the symmetry direction $x^{3}$.}
\label{fig:toroid}
\end{figure}

Our goal is to solve Eq.~(\ref{eq:curlcurl-1}) on a finite three-dimensional
domain $\Omega$ with a symmetry direction $\V{e}^{3} = \grad x^{3}$
along which the cross-section $\Omegat$ doesn't change and where all sources and boundary conditions are $2\pi$-periodic in $x^{3}$. An illustration in appropriately ordered cylindrical coordinates $x^{1} = Z$, $x^{2} = R$, $x^{3} = \varphi$ is found in Fig.~\ref{fig:toroid}.

\subsection{Vector calculus in 3D and 2D curvilinear coordinates}

Let $(x^{1}, x^{2}, x^{3})$ be right-handed curvilinear coordinates
that uniquely describe any position $\V{r} = \V{r}(x^{1}, x^{2}, x^{3})$
given by Cartesian components $r^{k}$ in the relevant domain. Co-
and contravariant basis vectors are respectively defined by 
\begin{equation}
\V{e}_{k} = \frac{\partial \V{r}}{\partial x^{k}} \quad \text{and} \quad\V{e}^{k} = \grad x^{k},
\end{equation}
and co- and contravariant components of vector fields $\V{V}(\V{r})$
by $V_{k} = \V{V} \cdot \V{e}_{k}$ and $V^{k} = \V{V} \cdot \V{e}^{k}$, respectively.
Components $g_{ij}$ of the metric tensor $\T{g}$ are
\begin{equation}
g_{kl} = \V{e}_{k} \cdot \V{e}_{l} = g_{kl}(x^{1}, x^{2}, x^{3}), \label{eq:gkl}
\end{equation}
and the metric determinant $g = \det [g_{kl}]$ is always positive,
as we are considering right-handed systems. Its square-root $\sqrt{g}$
is equal to the Jacobian of the transformation to Cartesian coordinates
and thus the weight of the volume element. The usual symmetry condition
$g_{kl} = g_{lk}$ follows from the geometric definition of $\T{g}$
in Eq.~(\ref{eq:gkl}). 

For the metric-free definition of differential operators and later
use in their discretised form for numerics it is useful to introduce
densities of weight $W$ to represent scalars, vectors, and tensors.
A density $\mathcal{U}$ (denoted in calligraphic letters) of weight
$W$ for a quantity $U$ is defined as
\begin{equation}
\mathcal{U} = \sqrt{g}^{W} U,
\end{equation}
where $U$ can represent either a scalar, or co/contravariant components
of a vector or tensor. If not stated otherwise, we use the term \emph{density}
for the default value $W=+1$. Usual scalars and vector/tensor components
correspond to the case $W=0$. Terms such as \emph{contravariant vector
density }and \emph{contravariant density representation of a vector
}will be used synonymously here for easier notation, as the conceptual
difference has no practical concequence in the present context.
\begin{table}
\caption{Natural input and output for differential
operators in curvilinear coordinates, densities are of weight $+1$
here.}
\label{tab:coordinates-1}
\centering\renewcommand{\arraystretch}{1.2}\renewcommand\cellgape{\Gape[3pt]}
\begin{tabular}{cccc}
\toprule
\textbf{Operator} & \textbf{Symbol} & \textbf{Input} & \textbf{Output}\tabularnewline
\midrule
Gradient & $\grad$ & scalar & covariant vector\tabularnewline
Curl & $\curl$ & covariant vector & \makecell{contravariant \\ vector density}\tabularnewline
Divergence & $\divg$ & \makecell{contravariant \\ vector density} & scalar density\tabularnewline
\midrule 
\makecell{Transverse \\ gradient} & $\gradt$ & scalar & covariant vector\tabularnewline
\makecell{Transverse \\ scalar curl} & $\scurlt$ & covariant vector & scalar density\tabularnewline
\makecell{Transverse \\ vector curl} & $\vcurlt$ & scalar & \makecell{contravariant \\ vector density}\tabularnewline
\makecell{Transverse \\ divergence} & $\divgt$ & \makecell{contravariant \\ vector density} & scalar density\tabularnewline
\bottomrule
\end{tabular}
\end{table}

\begin{table}
\caption{Translation between terminology of classical
tensor calculus and differential geometry in dimension $N$.}
\label{tab:coordinates-2}
\centering\renewcommand{\arraystretch}{1.2}
\begin{tabular}{cc}
\toprule
\textbf{Tensor calculus} & \textbf{Differential geometry}\tabularnewline
\midrule
scalar & $0$-form / scalar\tabularnewline
scalar density & $N$-form\tabularnewline
contravariant vector & vector\tabularnewline
covariant vector & $1$-form / covector\tabularnewline
contravariant vector density & $(N-1)$-form\tabularnewline
rank-2 tensor densities & Hodge operators\tabularnewline
\bottomrule 
\end{tabular}
\end{table}
Table~\ref{tab:coordinates-1} lists the choice of input and output
representation for differential operators such that the Jacobian $\sqrt{g}$
is formally removed from their definition. In this way, the coordinate-independent
definitions of 3D divergence and gradient are\footnote{Here and later we use the notation $\partial_{k}=\partial/\partial x^{k}$
and the usual convention to sum over indices appearing
twice, i.e. $\sum_{k=1}^{3}$ in Eq.~(\ref{eq:nabla-1}).}
\begin{equation}
\divg \V{U} = \partial_{k} \mathcal{U}^{k}, \quad \grad U= \V{e}^{k} \partial_{k} U, \label{eq:nabla-1}
\end{equation}
where the divergence acts on a contravariant vector density and yields
a scalar density, and the gradient acts on a scalar field and yields
a covariant vector field. The 3D curl operator
\begin{equation*}
\curl \V{U} = \epsilon^{ijk} \V{e}_{i} \partial_{j} U_{k}
\end{equation*}
acts on a covariant vector field and yields a contravariant vector
density. It contains the Levi-Civita tensor $\T{\epsilon}$ with contravariant
density components $\epsilon^{ijk} = 1$ for $ijk$ being circular
permutations of $123$, $-1$ for permutations of $321$, and $0$
otherwise. 

We want to treat the symmetry direction $x^3$ separately,
splitting the problem into a longitudinal part along the symmetry direction,
and a transverse part containing the remaining two dimensions.
For the notation of two-dimensional vectors we use lowercase bold letters,
i.e.
\begin{equation*}
\V{u} = U_{1} \V{e}^{1} + U_{2} \V{e}^{2} = U^{1} \V{e}_{1} + U^{2} \V{e}_{2}.
\end{equation*}
We define two-dimensional transverse divergence and gradient operators
analogous to the 3D case with 
\begin{equation}
\divgt \V{u} = \partial_{1} \mathcal{U}^{1} + \partial_{2} \mathcal{U}^{2}, \quad \gradt U = \V{e}^{1} \partial_{1} U + \V{e}^{2} \partial_{2} U. \label{eq:nabla-2}
\end{equation}
For the curl, in contrast to 3D, both a vectorial and a scalar transverse
curl operator exist with 
\begin{equation}
\vcurlt U \V{e}_{1} \partial_{2} U - \V{e}_{2} \partial_{1} U, \quad \V{u} = \partial_{1} U_{2} - \partial_{2} U_{1}, \label{eq:curl-1}
\end{equation}
yielding vector and scalar densities, respectively. Note that
all four of these operations either take a scalar corresponding to the longitudinal
part as input and give a two-dimensional vector corresponding to the transverse
part as output, or vice versa.

Matching input and output of operators in Table~\ref{tab:coordinates-1}
reflect the de Rham complex \citep{brezzi2012mixed} describing in
which order operators may act between different function spaces. In
3D this is
\begin{equation}
H^{1} \overset{\grad}{\rightarrow} H^{\curl} \overset{\curl}{\rightarrow} H^{\divg} \overset{\divg}{\rightarrow} L^{2}, \label{eq:derham}
\end{equation}
where application of two operators in a row yields zero, e.g.
$\curl \grad U = 0$. This is the case for any de Rham complex. The 3D de Rham diagram
leads from a scalar field in $H^{1}$ to a scalar density field in
$L^{2}$.

Since the curl operator mixes vector components and to distinguish
between transverse and longitudinal parts, relation~(\ref{eq:derham})
breaks up into two separate ones in 2D, given by
\begin{align}
H^{1} \overset{\gradt}{\rightarrow} H^{\scurlt} \overset{\scurlt}{\rightarrow} L^{2}, \label{eq:gradcurl}\\
H^{1} \overset{\vcurlt}{\rightarrow} H^{\divgt} \overset{\divgt}{\rightarrow} L^{2}. \label{eq:curldiv}
\end{align}
As in 3D, both diagrams lead from a scalar field in $H^{1}$ to a
scalar density field in $L^{2}$. The difference lies in the use of
covariant vectors in $H^{\scurlt}$
or contravariant vector densities in $H^{\divgt}$.
These two cases can be translated into each other by rotation via
the 2D Levi-Civita tensor $\lctenst$ with contravariant
density components given by
\begin{equation}
\epsilon_{\tv}^{kl} = \begin{pmatrix}
0 & 1 \\
-1 & 0
\end{pmatrix}. \label{eq:leviCivita}
\end{equation}
Relations between 2D differential operators can be written as
\begin{align}
\vcurlt U &= \lctenst \gradt U, \label{eq:grad-1} \\
\scurlt \V{u} &= \divgt (\lctenst \V{u}). \label{eq:div}
\end{align}

\subsection{Covariant formulation of classical electrodynamics}

Using differential operators in the metric-free way stated above,
Maxwell's equations act on density representations of fields listed
in Table~\ref{tab:fields}. In SI units they are written as
\begin{align}
\partial_{k} \mathcal{D}^{k} &= \varrho, \label{eq:Gauss} \\
\epsilon^{ijk} \partial_{j} E_{k} &= -\frac{\partial\mathcal{B}^{i}}{\partial t}, \label{eq:Faraday} \\
\epsilon^{ijk} \partial_{j} H_{k} &= \mathcal{J}^{i} + \frac{\partial\mathcal{D}^{i}}{\partial t}, \label{eq:AmpereMaxwell} \\
\partial_{k} \mathcal{B}^{k} &= 0, \label{eq:divB}
\end{align}
in any curvilinear coordinate system.
Magnetostatic equations (\ref{eq:magstat-1}-\ref{eq:divb-1})
are reproduced in the stationary limit of (\ref{eq:AmpereMaxwell}--\ref{eq:divB})
with vanishing time derivatives. The constitutive relations linking
excitations $\mathcal{D}^{k}, H_{k}$ to local linear responses $E_{l}, \mathcal{B}^{l}$
are 
\begin{align}
\mathcal{D}^{k} &= \varepsilon^{kl} E_{l}, \label{eq:defD} \\
H_{k} &= \nu_{kl} \mathcal{B}^{l}, \label{eq:defH}
\end{align}
with permittivity $\T{\varepsilon}$ and reluctivity $\T{\nu}$
respectively represented by contravariant $W=+1$ density and covariant
$W=-1$ density components,
\begin{align}
\varepsilon^{kl} &= \sqrt{g} \V{e}^{k} \cdot \T{\varepsilon} \V{e}^{l}, \label{eq:eps} \\
\nu_{kl} &= \sqrt{g}^{-1} \V{e}_{k} \cdot \T{\nu} \V{e}_{l}. \label{eq:nu}
\end{align}
While the field equations (\ref{eq:Gauss}--\ref{eq:divB}) remain
coordinate-independent, Eqs.~(\ref{eq:defD}--\ref{eq:defH}) contain all influence from the
metric tensor $\T{g}$ implicitly via the basis vectors and the Jacobian in
Eqs.~(\ref{eq:eps}--\ref{eq:nu}). In particular for a scalar $\nu$, components
of $\T{g}$ enter the resulting covariant density representation
of $\T{\nu}$ in curvilinear  coordinates,
\begin{equation}
\nu_{kl} = \frac{1}{\sqrt{g}} \sum_{i, j} \partial_{k} r^{i} \partial_{l} r^{j} \nu \delta_{ij} = \frac{g_{kl}}{\sqrt{g}} \nu. \label{eq:scalarnu}
\end{equation}
This means that covariant reluctivity components generated by a
scalar $\nu$ inherit their symmetry properties from $g_{kl}$. If
physical components of the permeability tensor are already given in
the desired coordinate frame as a matrix $[\mu_{(kl)}]$, covariant
density components of $\T{\nu}$ can be found by taking its inverse
$[\nu_{(kl)}]=[\mu_{(kl)}]^{-1}$ and computing 
\begin{equation}
\nu_{kl} = \frac{\sqrt{g_{kk} g_{ll}}}{\sqrt{g}} \nu_{(kl)}.
\end{equation}
One can see that if $\nu_{kl}$ are constant in certain curvilinear
coordinates, physical components $\nu_{(kl)}$ will usually vary locally
and vice-versa. One should remark that the solution for covariant
components $H_{k}$ in curvilinear geometry with constant physical
$\nu_{(kl)}$ is identical to one for Cartesian components of $\V{H}$
with a spatially varying $\nu_{(kl)}$. Thus one could emulate curvilinear
geometry for magnetostatics in flat geometry by a material with locally
varying permeability, and vice versa.

\begin{table}
\caption{Conventions to represent electromagnetic scalar,
vector and tensor fields by densities of varying weight. }
\label{tab:fields}
\centering\renewcommand{\arraystretch}{1.2}
\begin{tabular}{cccc}
\toprule
\textbf{Quantity} & \textbf{Symbol} & \textbf{Representation} & \textbf{Weight}\tabularnewline
\midrule
Metric tensor & $g_{kl}$ & covariant & $0$\tabularnewline
Inverse metric & $g^{kl}$ & contravariant & $0$\tabularnewline
Jacobian & $\sqrt{g}$ & scalar & $+1$\tabularnewline
Levi-Civita tensor & $\epsilon^{ijk}$ & contravariant & $+1$\tabularnewline
Charge density & $\varrho$ & scalar & $+1$\tabularnewline
Current density & $\mathcal{J}^{k}$ & contravariant & $+1$\tabularnewline
Electric field & $E_{k}$ & covariant & $0$\tabularnewline
Magnetic flux density & $\mathcal{B}^{k}$ & contravariant & $+1$\tabularnewline
Electric displacement & $\mathcal{D}^{k}$ & contravariant & $+1$\tabularnewline
Magnetic field & $H_{k}$ & covariant & $0$\tabularnewline
Scalar potential & $\Phi$ & scalar & $0$\tabularnewline
Vector potential & $A_{k}$ & covariant & $0$\tabularnewline
Permittivity & $\varepsilon^{kl}$ & contravariant & $+1$\tabularnewline
Permeabiliy & $\mu^{kl}$ & contravariant & $+1$\tabularnewline
Reluctivity & $\nu_{kl}$ & covariant & $-1$\tabularnewline
\midrule
Longitudinal reluctivity & $\nu_{33}$ & scalar & $-1$\tabularnewline
Transverse reluctivity & $\nu_{\tv,\,kl}$ & covariant & $-1$\tabularnewline
Mod. transv. reluctivity & $\modimpt^{kl}$ & contravariant & $+1$\tabularnewline
Transverse Levi-Civita tensor & $\epsilon_{\tv}^{kl}$ & contravariant & $+1$\tabularnewline
\bottomrule
\end{tabular}
\end{table}

\subsection{Reduction of magnetostatics to 2D by Fourier expansion}

To reduce the 3D problem of Eq.~(\ref{eq:curlcurl-1}) to a number
of 2D equations we write quanties assumed to be $2\pi$-periodic in
$x^{3}$ as a Fourier series 
\begin{equation}
f(x^{1}, x^{2}, x^{3}) = \sum_{n = -\infty}^{\infty} f_{n}(x^{1}, x^{2}) \, e^{\imun nx^{3}}.
\end{equation}
The following equations concern single harmonics with $n$ omitted
as an index in the notation. To be compatible with the curl-curl equation
(\ref{eq:curlcurl-1}) in covariant form we require a Fourier expansion
of \emph{covariant vector} components $A_{k}$ of $\V{A}$ and \emph{contravariant
vector density} components $\mathcal{J}^{k}$ of $\V{J}$. 

To retain linearity of terms involving $x^{3}$
and thus avoid mode-coupling via convolution in the constitutive
relation~(\ref{eq:defH}), we require covariant $W=-1$ density components
$\nu_{kl} = \nu_{kl}(x^{1}, x^{2})$ of the impermebility tensor $\T{\nu}$
to be independent of the symmetry coordinate $x^{3}$ if we allow
arbitrary harmonics $n$ for the fields. In addition, to be able to
split transverse and longitudinal components of fields later, off-diagonal
components in $x^{3}$ shall vanish. This means that 
\begin{equation}
\nu_{kl} = \begin{pmatrix}
\nu_{\tv,\,kl} & \begin{array}{c} 0 \\ 0 \end{array} \\
\begin{array}{cc} 0 & 0 \end{array} & \nu_{33}
\end{pmatrix}, \label{eq:nublock}
\end{equation}
with transverse reluctivity
\begin{equation}
\nu_{\tv,\,kl} = \begin{pmatrix}
\nu_{11} & \nu_{12} \\
\nu_{21} & \nu_{22}
\end{pmatrix} \label{eq:cov}
\end{equation}
as a covariant $W=-1$ density and longitudinal reluctivity $\nu_{33}$
as a scalar $W=-1$ density. In the case of a scalar permeability,
those symmetry restrictions thus apply to the metric tensor and vice
versa, so $\sqrt{g}^{-1} g_{kl}$ shall be independent from $x^{3}$
and $g_{k3} = g_{3k} = 0$ for $k \neq 3$. A modified transverse reluctivity
\begin{equation}
\modimptenst = -\lctenst \T{\nu}_{\tv} \lctenst \label{eq:nu-1}
\end{equation}
is useful to introduce, with contravariant density components 
\begin{equation*}
\modimpt^{kl} = \begin{pmatrix}
\nu_{22} & -\nu_{21} \\
-\nu_{12} & \nu_{11}
\end{pmatrix},
\end{equation*}
proportional to the transpose of the inverse of (\ref{eq:cov}). For
a scalar $\nu$ and $g_{kl}$ of the form (\ref{eq:nublock}) this
reduces to 
\begin{equation}
\modimpt^{kl} = \frac{\sqrt{g}\,g_{\tv}^{kl}}{g_{33}} \nu,
\end{equation}
where $g_{\tv}^{kl}$ is the transverse part of the (symmetric) inverse
metric tensor.\footnote{If instead the inverse of the transverse part of $\T{g}$ were used,
the result would be identical, e.g. for cylindrical coordinates
$\modimpt^{ZZ} = \modimpt^{RR} = \nu/R$. This inverted dependency
on $R$ compared to the usual Laplacian $\upDelta$ in cylindrical coordinates
is characteristic for the Grad-Shafranov operator $\upDelta^{\star}$
(see \citep{Dhaeseleer_BOOK_1991}).}

Under the given restrictions the non-oscillatory part $n = 0$ of Eq.~(\ref{eq:curlcurl-1})
splits into a transverse and a longitudinal part with 
\begin{align}
\vcurlt \nu_{33} \scurlt \V{a} &= \V{j}, \label{eq:axitrans} \\
\scurlt \T{\nu}_{\tv} \vcurlt A_{3} &= \mathcal{J}^{3}, \label{eq:J3}
\end{align}
where two-component $\V{a} = A_{1} \V{e}^{1} + A_{2} \V{e}^{2}$
are expressed as a covariant vector and $\V{j} = \mathcal{J}^{1} \V{e}_{1} + \mathcal{J}^{2} \V{e}_{2}$
as a contravariant vector density. The labels transverse for Eq.~(\ref{eq:axitrans})
and longitudinal for Eq.~(\ref{eq:J3}) refer to $\V{A}$
and $\V{J}$ here. This is not to be confused with $\V{B}$
itself, for which exactly the opposite is the case, i.e. the longitudinal
$\mathcal{B}^{3}$ follows from Eq.~(\ref{eq:axitrans}) and the
transverse $\V{b}$ from Eq.~(\ref{eq:J3}), corresponding to the first
step in Eq.~(\ref{eq:gradcurl}) and (\ref{eq:curldiv}), respectively. Using the
relations between 2D differential operators in Eqs.~(\ref{eq:grad-1}-\ref{eq:div})
together with the modified transverse reluctivity $\modimptenst$
of Eq.~(\ref{eq:nu-1}), we can re-write Eq.~(\ref{eq:J3}) as
\begin{equation}
-\divgt (\modimptenst \gradt A_{3}) = \mathcal{J}^{3}. \label{eq:longi-1}
\end{equation}
For the oscillatory part with $n \neq 0$, we set the covariant component
$A_{3}$ to zero by a gauge transformation 
\begin{equation}
\V{A} \rightarrow \V{A} - \grad \int_{x_{0}^{3}}^{x^{3}} A_{3} \, \diff x^{3 \prime}. \label{eq:Agauge}
\end{equation}
This corresponds to setting $\V{A} \rightarrow \V{A} - \grad A_{3}/(\imun n)$
for single harmonic $\V{A} = \V A_{n}(x_{1}, x_{2}) e^{\imun nx^{3}}$ with
harmonic index $n$. Independent gauging in that manner is possible
for each $n \neq 0$. Due to the superposition principle, we can use
a fully transverse vector potential with a different gauge for each
individual harmonic and take a Fourier sum in the end. In that case
the contravariant magnetic flux density components are given by
\begin{equation}
\mathcal{B}^{1} = -\imun n A_{2}, \quad \mathcal{B}^{2} = \imun n A_{1}, 
\quad \mathcal{B}^{3} = \scurlt \V{a}.
\end{equation}
Ampère's law is transformed to
\begin{align}
\partial_{2} (\nu_{33} \scurlt \V{a}) + n^{2} (\nu_{22} A_{1} - \nu_{21} A_{2}) &= \mathcal{J}^{1}, \label{eq:amp1} \\
-\partial_{1} (\nu_{33} \scurlt \V{a}) + n^{2} (-\nu_{12} A_{1} + \nu_{11} A_{2}) &= \mathcal{J}^{2}, \label{eq:amp2} \\
%
%
\imun n \left[\partial_{1} (\nu_{22} A_{1} - \nu_{21} A_{2}) 
            + \partial_{2} (\nu_{11} A_{2} - \nu_{12} A_{1})\right] &= \mathcal{J}^{3}. \label{eq:amp3}
\end{align}
Eqs.~(\ref{eq:amp1}-\ref{eq:amp2}) contain only transverse
components of $\V{A}$ and $\V{J}$ and can be rewritten by using the
vector $\vcurlt$ operator, yielding the purely
2D problem for $n \neq 0$ as
\begin{equation}
\vcurlt \left(\nu_{33} \scurlt \V{a}\right) + n^{2} \modimptenst \V{a} = \V{j}. \label{eq:curlcurl-2}
\end{equation}
As opposed to the singular ungauged three-dimensional \textbf{curl}-\textbf{curl}
equation (\ref{eq:axitrans}), the additional term resulting from
the fixed gauge makes Eq. (\ref{eq:curlcurl-2}) uniquely solveable,
analogous to the 3D \textbf{curl}-\textbf{curl} equation with a time-harmonic
term. Like in the scalar Helmholtz equation $-\upDelta \Phi + n^{2} \Phi = \rho$
arising from Fourier expansion of the Poisson equation, the positive
sign of the term weighted by $n^{2}$ leads to rapidly decaying solutions,
opposed to oscillating solutions which would typically result from
the Fourier expansion of a wave equation in time. Eq.~(\ref{eq:amp3})
links longitudinal $\boldsymbol{J}$ and $\boldsymbol{B}$ components and can be written compactly as
\begin{equation}
\imun n \divgt (\modimptenst \V{a}) = \mathcal{J}^{3}. \label{eq:J3n}
\end{equation}
Eq.~(\ref{eq:J3n}) is automatically fulfilled via the divergence relation for Fourier
harmonics
\begin{equation}
\divgt \V{j} + \imun n \mathcal{J}^{3} = 0, \label{eq:divJ}
\end{equation}
which can be seen from applying $\divgt$ to Eq.~(\ref{eq:curlcurl-2}).

Formally, by setting $n = 0$, Eq.~(\ref{eq:curlcurl-2}) includes
Eq.~(\ref{eq:axitrans}) as a special case, so we summarize the two
as the class of \emph{transverse} equations (\ref{eq:curlcurl-2})
for arbitrary $n$, whereas the \emph{longitudinal} equation~(\ref{eq:longi-1})
equivalent to Eq.~(\ref{eq:J3}) needs only to be solved for $n = 0$.

\subsection{Variational formulation in coordinate space}

To find a variational formulation for numerical computations, Eqs.~(\ref{eq:longi-1})
and (\ref{eq:curlcurl-2}) are multiplied by a test function and integrated
with weight $\sqrt{g}$ in coordinate space $(x^{1}, x^{2})$ parametrising
the cross-section perpendicular to the symmetry direction of the original
domain $\Omega$ with a 2D volume element $\diff \Omegat := \diff x^{1} \, \diff x^{2}$
and line element on the boundary, 
\begin{equation}
\diff \Gammat = \sqrt{\left(\diff x^{1}\right)^{2} + \left(\diff x^{2}\right)^{2}}.
\end{equation}
Volume integration of quantities $F(x^{1}, x^{2})$ over $\Omega$
and dividing the result by the range $2\pi$ of $x^{3}$ results in
an integral over $\Omega_{\text{t}}$ of the density representation
$\mathcal{F} = \sqrt{g} F$,
\begin{equation}
\begin{split}
\frac{1}{2\pi} \int_{\Omega} F(x^{1}, x^{2}) \, \diff V = \frac{1}{2\pi} \int_{0}^{2\pi} \diff x^{3} \int_{\Omegat} \diff x^{1} \, \diff x^{2} \, \mathcal{F}(x^{1}, x^{2}) \twocolbreak = \int_{\Omegat} \mathcal{F}(x^{1}, x^{2}) \, \diff \Omegat.
\end{split}
\end{equation}
The variational form of the longitudinal equation~(\ref{eq:longi-1})
for the non-oscillatory part $n = 0$ with scalar test function $w(x^1, x^2)$ is
\begin{equation}
\begin{split}
\int_{\Omegat} (\partial_{k} w) \modimpt^{kl} (\partial_{l} A_{3}) \, \diff \Omegat \twocolbreak - \int_{\Gammat} w n_{k} \modimpt^{kl} (\partial_{l} A_{3}) \, \diff \Gammat = \int_{\Omegat} w \mathcal{J}^{3} \, \diff \Omegat. \label{eq:poisson-1}
\end{split}
\end{equation}
Here $\V{n} = (n_{1}, n_{2})$ is the unit outward normal vector across
the boundary line $\Gammat$ in coordinate space $(x^{1}, x^{2})$
and the implied sums are taken over $k, l$ from $1$ to $2$. For the special case
$w = 1$ a compatibility condition follows as
\begin{equation}
-\int_{\Gammat} n_{k} \modimpt^{kl} \partial_{l} A_{3} \, \diff \Gammat = \int \mathcal{J}^{3} \, \diff \Omegat,
\end{equation}
fixing the Neumann term in Eq.~(\ref{eq:poisson-1}) corresponding
to the magnetic field parallel to the transverse boundary to the total
current through the surface $x^{3} = \mathrm{const.}$ within the domain.

For the transverse equation (\ref{eq:curlcurl-2}), with vectorial test function $\V{w}$
and denoting the coordinate vector along the boundary in counter-clockwise direction by
$\V{s} = (s^{1}, s^{2}) = -\lctenst \V{n} = (-n_{2}, n_{1})$, we obtain
\begin{equation}
\begin{split}
\int_{\Omegat} \scurlt \V{w} \, \nu_{33} \scurlt \V{a} \, \diff \Omegat + n^{2} \int_{\Omegat} w_{k} \modimpt^{kl} A_{l} \, \diff \Omegat \twocolbreak - \int_{\Gammat} w_{k} s^{k} \nu_{33} \scurlt \V{a} \, \diff \Gammat = \int_{\Omegat} w_{k} \mathcal{J}^{k} \, \diff \Omegat, \label{eq:weakneumann}
\end{split}
\end{equation}
for the case $n = 0$ as well as $n \neq 0$.

\subsection{Cartesian and cylindrical coordinates}

For Cartesian coordinates and scalar reluctivity $\nu$, the variational form of
Eq.~(\ref{eq:weakneumann}) with Neumann boundary condition on $\Gammat$
and test function $\V{w}$ is given by
\begin{equation}
\begin{split}
\int_{\Omegat} \scurlt \V{w} \, \nu \scurlt \V{a} \, \diff x \, \diff y + n^{2} \int_{\Omegat} \V{w} \cdot \nu \V{a} \, \diff x \, \diff y \twocolbreak{} - \int_{\Gammat} \V{s} \cdot \V{w} \nu \scurlt \V{a} \, \diff \Gamma = \int_{\Omegat} \V{w} \cdot \V{j} \, \diff x \, \diff y. \label{eq:weakform}
\end{split}
\end{equation}
Here, $\V{s} = (-n_{y}, n_{x})$ is the tangential vector along the boundary
line $\Gammat$ of the $xy$ cut of the domain, $\Omegat$,
and $\V{n} = (n_{x}, n_{y})$ the unit outward normal vector orthogonal
to $\V{s}$ in this cross-section. 

For cylindrical coordinates\footnote{The re-ordering from $(R, \varphi, Z)$ to $(Z, R, \varphi)$ is required
to fit the general framework with symmetry coordinate $x^{3}$.} $x^{1} = Z, \, x^{2} = R, \, x^{3} = \varphi$, the non-vanishing metric components
are $g_{11} = g_{22} = 1$ and $g_{33} = R^{2}$ and Eq. (\ref{eq:weakneumann})
with a scalar permeability $\nu$ yields
\begin{equation}
\begin{split}
\int_{\Omegat} (\partial_{Z} w_{R} - \partial_{R} w_{Z}) R \nu (\partial_{Z} A_{R} - \partial_{R} A_{Z}) \, \diff R \, \diff Z \twocolbreak
+ n^{2} \int_{\Omegat} \frac{\nu}{R} (w_{R} A_{R} + w_{Z} A_{Z}) \, \diff R \, \diff Z \\
- \int_{\Gammat} (w_{R} s^{R} + w_{Z} s^{Z}) R \nu (\partial_{Z} A_{R} - \partial_{R} A_{Z}) \, \diff \Gammat \twocolbreak
= \int_{\Omegat} (w_{R} \mathcal{J}^{R} + w_{Z} \mathcal{J}^{Z}) \, \diff R \, \diff Z. \label{eq:weakneumann-1}
\end{split}
\end{equation}

\section{Finite element discretisation}

For the general transverse variational problem in Eq.~(\ref{eq:weakneumann}),
the natural discretisation for $\V{a}$ are 2D (Nédélec)
edge elements conforming to $H(\curl, \Omegat)$. Due
to its fixed divergence in Eq.~(\ref{eq:J3}), components $\mathcal{J}^{k}$
of the transverse current density $\V{j}$ should be discretized
by 2D (Raviart-Thomas) elements conforming to $H(\divg, \Omegat)$.
For practical reasons, it is convenient to define a quantity $\V{t}$
in $H(\curl, \Omegat)$ with $\divgt \V{j} = \scurlt \V{t}$
instead. Comparison of components yields
\begin{equation}
T_{1} = -\mathcal{J}^{2}, \quad T_{2} = \mathcal{J}^{1},
\end{equation}
which can also be expressed via the Levi-Civita tensor defined in
Eq.~(\ref{eq:leviCivita}) as 
\begin{equation}
\V{j} = \lctenst \V{t}.
\end{equation}

The 2D vector $\V{t}$ is related to the $n$th harmonic
of the current vector potential $\V{T} \propto e^{\imun nx^{3}}$
with $\V{J} = \vcurlt \V{T}$, what
can be seen from the corresponding expressions. As with Eq.~(\ref{eq:Agauge}),
a fully transverse $\V{T}$ for $n \neq 0$ can be chosen as
$\V{T} = T_{1} \V{e}^{1} + T_{2} \V{e}^{2}$,
to which $\V{t}$ of Eq.~(\ref{eq:leviCivita}) is proportional with factor
$\imun n$. For $n = 0$, one can instead choose a longitudinal $\V{T} = T \V{e}^{3}$
with scalar stream function $T(x^{1}, x^{2})$ and $\V{j} = \vcurlt T$,
as long as $\mathcal{J}^{3} = 0$.

\section{Validation results}

Here we present results for an implementation of the presented Fourier-FEM
approach in cylindrical coordinates. To discretize the variational
formulation, 2D Raviart-Thomas (longitudinal) and Nédélec (transversal)
finite elements of lowest order are used inside FreeFEM~\citep{FreeFEM}.
Validity and performance of numerical compuations are assessed based
on analytical test cases with field components defined in a piecewise
manner over cylinder radius $R$. For the numerical computations only
boundary values and possible volumetric currents are imposed.

\begin{figure}
\centering
\includegraphics{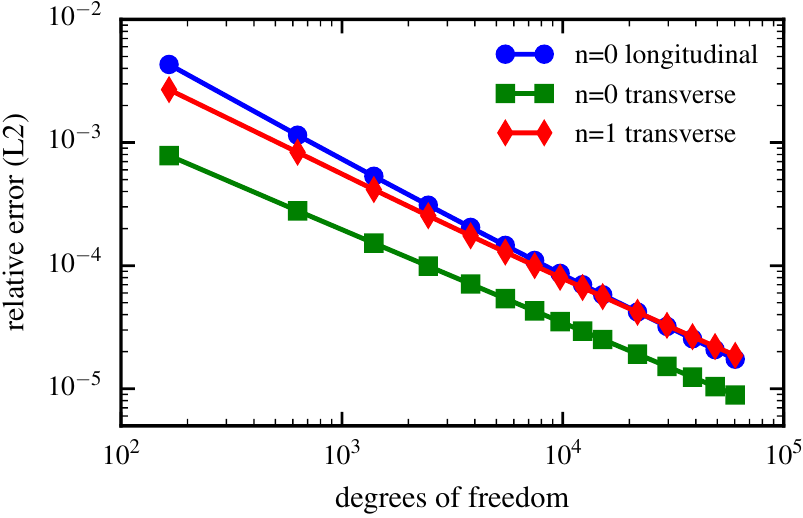}
\caption{Convergence of numerical Fourier-FEM computations for analytical test
cases. As expected the relative error decreases linearly with the
number of degrees of freedom.}
\label{fig:convergence}
\end{figure}

For the axisymmetric ($n = 0$) case we introduce an analytical longitudinal
magnetic field $\V{B} \parallel \V{e}^{Z}$ with
\begin{align}
\mathcal{B}^{Z} &= \begin{cases}
\frac{52}{25} R - \frac{R^{3}}{2} & (R < 0.4), \\
100R & (0.4 \leq R < 0.5),\\
2R & (R \geq 0.5),
\end{cases}
\end{align}
resulting in
\begin{equation}
\mathcal{J}^{R} = 0, \quad \mathcal{J}^{Z} = 0,
\end{equation}
and
\begin{equation}
\mathcal{J}^{\varphi} = J_{(\varphi)} = \begin{cases}
1 & (R < 0.4), \\
0 & (R \geq 0.4),
\end{cases}
\end{equation}
and a second case with a transverse field $\V{B}\parallel\V{e}^{\varphi}$
given by 
\begin{equation}
\mathcal{B}^{\varphi} = \begin{cases}
\frac{R}{2} & (R < 0.4), \\
\frac{4}{R} & (0.4 \leq R < 0.5), \\
\frac{2}{25R} & (R \geq 0.5),
\end{cases}
\end{equation}
resulting in
\begin{equation}
\mathcal{J}^{R} = 0, \quad \mathcal{J}^{\varphi} = 0,
\end{equation}
and
\begin{equation}
\mathcal{J}^{Z} = RJ_{(Z)} = \begin{cases}
R & (R < 0.4), \\
0 & (R \geq 0.4).
\end{cases}
\end{equation}
As an analytical test case for the non-axisymmetric Fourier harmonic
$n = 1$, we use a radial transverse field with $\V{J} = \V{0}$
and
\begin{equation}
\mathcal{B}_{n}^{R} = \begin{cases}
\frac{20000}{128927} R & (R < 0.4), \\
\frac{510000}{128927} R - \frac{78400}{128927} \frac{1}{R} & (0.4 \leq R < 0.5), \\
\frac{106436}{128927} R + \frac{22491}{128927} \frac{1}{R} & (R \geq 0.5).
\end{cases}
\end{equation}
Fig.~\ref{fig:convergence} shows the convergence of numerical computations.
The convergence rate of the L2 error over degrees of freedom is linear,
as expected from the discretization in the lowest order edge element
space.

\section{Conclusion and Outlook}

An approach for the numerical solution for the magnetostatic field
on 3D domains with a symmetry direction has been described. In particular
it is applicable to translationally symmetric and axisymmetric domains.
Its validity has been demonstrated on model problems in cylindrical
coordinatesSince each harmonic is computed separately, batch parallel
computations are easily possible using the described method.

Even though arbitrary spatial variations of current density and boundary
conditions can be treated, the approach is most efficient for symmetric
current distributions reducing the number of non-zero harmonics. In
axisymmetric domains this includes circular ($n = 0$) or square-like
shapes ($n = 0, 4, 8, \dotsc$). Furthermore, smoother dependencies on the
symmetry coordinate lead to a faster decay in the spectrum. A more
severe limitation for engineering applications, apparent in the chosen
model problems, is the restriction on the shape of regions with different
permeability following the symmetry direction. To lift this requirement,
coupling of different harmonics would have to be taken into account,
thereby removing the original advantage of trivial parallelisability.

The method may be generalised further by considering an expansion
in a different set of functions, e.g. Bessel functions for the expansion
in cylindrical harmonics in the radial direction. For the treatment
of time-dependent and time-harmonic eddy current or full electromagnetic
wave problems the introduction of a an additional electromagnetic
scalar potential will be required for the oscillatory harmonics if
the gauge is fixed to eliminate one component of the vector potential.

\section*{Acknowledgements}

The authors would like to thank Fabian Weissenbacher for supporting
initial investigations. Special thanks go to Friedrich Hehl for insightful
discussions on the covariant formulation of electromagnetism. The
authors gratefully acknowledge support from NAWI Graz, from the OeAD
under the WTZ grant agreement with Ukraine No UA 04/2017 and from the Reduced Complexity Models grant number ZT-I-0010 by the Helmholtz Association of German Research Centers.

\bibliographystyle{IEEEtran}
\bibliography{paper_magnetic}

\end{document}